\documentclass[aps,prl,amsmath,nofootinbib,twocolumn,10pt,showpacs,preprintnumbers,superscriptaddress]{revtex4-1}

\usepackage{mystyle}
\usepackage{pgfkeys}
\usepackage{xcolor}

\usepackage[utf8]{inputenc} 

\newcommand{\hybridmaxuferr}{1\times 10^{-4}}

\newcommand{\optional}[1]{}

\newcommand{\code}[1]{{\tt #1}}

\newcommand{\Hz}{\text{ Hz}}

\newcommand{\doubleline}{\hline \hline}

\usepackage{printlen}

\begin{document}

\title{Gravitational waveforms for high spin and high mass-ratio
  binary black holes:\\ A synergistic use of numerical-relativity codes}

\newcommand{\CITA}{\affiliation{Canadian Institute for Theoretical Astrophysics, University of Toronto, Toronto M5S 3H8, Canada}}
\newcommand{\Cornell}{\affiliation{Cornell Center for Astrophysics and Planetary Science, Cornell University, Ithaca, New York, 14853, USA}}
\newcommand{\Maryland}{\affiliation{Department of Physics, University of Maryland, College Park, MD 20742, USA}}

\newcommand{\AEI}{\affiliation{
Max Planck Institute for Gravitational Physics (Albert Einstein Institute),
Am M\"uhlenberg 1, Potsdam 14476, Germany}
}

\author{Ian Hinder}
\email{ian.hinder@aei.mpg.de}
\AEI

\author{Serguei Ossokine}
\email{serguei.ossokine@aei.mpg.de}
\AEI

\author{Harald P. Pfeiffer}
\email{harald.pfeiffer@aei.mpg.de}
\AEI \CITA 

\author{Alessandra Buonanno}
\email{alessandra.buonanno@aei.mpg.de}
\AEI \Maryland
\date{\today}

\begin{abstract}
Observation and characterisation of gravitational waves from binary
black holes requires accurate knowledge of the expected waveforms.
The late inspiral and merger phase of the waveform is obtained through
direct numerical integration of the full 3-dimensional Einstein
equations.  The Spectral Einstein Code (\code{SpEC}) utilizes a
multi-domain pseudo-spectral method tightly adapted to the geometry of
the black holes; it is computationally efficient and accurate, but---
for high mass-ratios and large spins---sometimes requires manual
fine-tuning for the merger-phase of binaries.  The Einstein Toolkit
(\code{ET}) employs finite difference methods and the moving puncture
technique; it is less computationally efficient, but highly robust.
For some mergers with high mass ratio and large spins, the efficient
numerical algorithms used in SpEC have failed, whereas the simpler
algorithms used in the \code{ET} were successful.  Given the urgent need of
testing the accuracy of waveform models currently used in LIGO and
Virgo inference analyses for high mass ratios and spins, we present
here a synergistic approach to numerical-relativity: We combine
\code{SpEC} and \code{ET} waveforms into complete inspiral-merger-ringdown waveforms, taking advantage
of the computational efficiency of the pseudo-spectral code during the
inspiral, and the robustness of the finite-difference code at the
merger.  We validate our method against a case where complete
waveforms from both codes are available, compute three new hybrid
numerical-relativity waveforms, and compare them with analytical
waveform models currently used in LIGO and Virgo science.  All the waveforms
and the hybridization code are publicly available.

\end{abstract}

\maketitle{}

{\it Introduction.}
In 2015,
Advanced LIGO~\cite{TheLIGOScientific:2014jea,TheLIGOScientific:2016agk}
detected the gravitational wave (GW) event GW150914 corresponding to
the merger of a binary black hole (BBH) system~\cite{Abbott:2016blz}.  Subsequently,
five further GW observations from BBH and binary neutron-star
mergers have been reported~\cite{Abbott:2016nmj,Abbott:2017vtc,Abbott:2017oio,TheLIGOScientific:2017qsa,Abbott:2017gyy}, with the Advanced Virgo
detector~\cite{TheVirgo:2014hva} participating in the more recent observations~\cite{Abbott:2017oio,TheLIGOScientific:2017qsa}. The parameters of the BBHs
were inferred~\cite{TheLIGOScientific:2016wfe,Abbott:2016izl,Abbott:2016nmj,Abbott:2017vtc,Abbott:2017oio,TheLIGOScientific:2017qsa,Abbott:2017gyy}
using fast, analytical waveform
models~\cite{Pan:2013rra,Hannam:2013oca,Taracchini:2013rva,Khan:2015jqa,Bohe:2016gbl,Babak:2016tgq} calibrated to numerical
relativity (NR) simulations~\cite{HannamEtal:2010,Buchman:2012dw,Mroue:2013xna,Husa:2015iqa,Kumar:2016dhh}.  The latter are obtained by directly solving the fully nonlinear
Einstein equations for BBH spacetimes~\cite{Baumgarte-Shapiro-Book}.  The detection
efficiency and the accuracy of parameter estimation are directly
affected by the accuracy of both analytical and numerical waveform models~\cite{Abbott:2016wiq}. 

The first complete BBH simulation was obtained
by Pretorius in
2005~\cite{Pretorius:2005gq} using finite difference methods and BH excision.  The finite difference moving puncture method, which
does not require a complex excision algorithm, was developed
independently by two groups in late 2005
\cite{Baker:2005vv,Campanelli:2005dd}; because of its simplicity and robustness, this method is in widespread use.  Finally, the Spectral Einstein 
Code~\cite{SpECwebsite} (\code{SpEC}) makes
use of accurate pseudo-spectral collocation methods combined with BH
excision.
 \code{SpEC}
improved on existing finite difference solutions in accuracy by one to
two orders of magnitude~\cite{Hannam:2009hh,Hinder:2013oqa} and roughly doubled the number of GW cycles in large waveform catalogs~\cite{Mroue:2013xna,Kumar:2016dhh}. \texttt{SpEC} has performed the longest BBH simulation to date (350GW cycles)~\cite{Szilagyi:2015rwa}, as well as the simulations with the largest BH spins (spin-magnitudes 0.998)~\cite{Chatziioannou:2018wqx}.  The largest mass rato simulations, however, have been performed using moving puncture finite difference codes~\cite{Lousto:2010ut,Husa:2015iqa}.

Numerical simulations are computationally expensive, typically
running for months on large supercomputers.
The computational cost to achieve a given accuracy increases with (i)
number of orbits $N_\text{orbs}$, (ii) the ratio of the masses of the
binaries $q = m_1/m_2 \ge 1$, and (iii) increasing magnitudes of the BH spins.
So far, analytic waveform models for BBHs used in LIGO and Virgo follow-up
analyses~\cite{Pan:2013rra,Taracchini:2013rva,Hannam:2013oca,Khan:2015jqa,Bohe:2016gbl,Babak:2016tgq}
have been calibrated to NR waveforms for mild mass ratios and spin magnitudes (e.g.,
Fig.~1 in~\cite{Bohe:2016gbl}),
using NR waveforms of $\sim 8\mbox{--}30$ orbits, depending
on the binary's parameters and the NR code.  The most extreme mass-ratio simulations with spins were presented in~\cite{Husa:2015iqa}. Validation of
waveform models for the earlier inspiral and for BBH parameters too extreme for
NR simulations have relied on internal consistency tests (%
e.g., Ref.~\cite{Pan:2013tva} for nonspinning BBHs), comparison of two-body dynamics in absence of NR-calibration \cite{Damour:2011fu,LeTiec:2011bk,LeTiec:2011dp,Ossokine:2017dge}, and comparisons between
NR-calibrated waveform models
 with common assumptions and input data\footnote{
  Certain NR waveforms are used in the calibration of both waveform models from
  Refs.~\cite{Khan:2015jqa,Bohe:2016gbl}; moreover, an uncalibrated effective-one-body
  model~\cite{Taracchini:2013rva}, which differs from the one underlying {\tt SEOBNRv4} for a few higher-order PN terms, is used during the construction of {\tt IMRPhenomD}.}.
Thus, the true accuracy of waveform models 
in regions of the parameter space without NR simulations is unknown.  

 Particularly important are mass ratios $q\!\gtrsim\! 4$, and
 BH with a large spin-projection along the orbital angular momentum,
 $\chi_{1,2}\equiv
 (\mathbf{S}_{1,2}\cdot\hat{\mathbf{L}})/m_{1,2}^2\gtrsim 0.8$
(e.g., Fig.~1 in~\cite{Bohe:2016gbl}).
These cases reach particularly high orbital frequency near merger
 and waveform models needed~\cite{Kumar:2016dhh,Bohe:2016gbl} (or will need~\cite{Nagar:2018zoe}) 
 recalibration to new NR waveforms when these became
 Only accurate NR waveforms in this region of parameter space
   will ensure that GW parameter measurements in upcoming LIGO and
   Virgo observations are free of biases due to errors in the waveform
   models.

Unfortunately,
this region of parameter space is
computationally very
challenging.
Resolving the multiple length and time scales associated with both the
large and small BH --- made even smaller and more distorted by its
high spin---
requires high resolution with consequent high computational cost.
BH horizons near merger become also particularly distorted
  complicating BH excision in the \texttt{SpEC} code, so that BBH
  inspiral simulations with spins $\chi_1\!=\!\chi_2\!=\!0.9$ at mass-ratios
  $q=3,4,5$ through merger fail shortly before merger.

In contrast, finite difference codes are able to evolve BBHs with more
extreme mass-ratios and spins through merger, albeit only starting
with a higher orbital frequency (i.e.~lower separation, and covering fewer
orbits).

The higher initial frequency implies
  a significantly higher minimum total mass, $M = m_1+m_2$,
for which the waveform still covers the entire LIGO frequency band.
While these
shorter waveforms allow calibrating waveform
models near merger, the extra low frequency information of
longer SpEC simulations may be essential to achieve
faithful models and avoid biases in parameter estimation and tests of general relativity from detected
GW signals.

Whilst we are confident that the problems with \code{SpEC} will be solved by improved numerical
methods, the upcoming LIGO and Virgo runs, which promise several tens of 
GW events from BBHs~\cite{Aasi:2013wya}, increase the urgency for accurate waveforms from this region
of the parameter space. Thus, we explore here a synergistic approach that combines highly accurate \code{SpEC}
inspiral evolutions with shorter merger waveforms from finite
difference codes (in this case, the Einstein Toolkit~\cite{Loffler:2011ay,etweb}) to produce
\textit{NR-NR hybrid} waveforms. In this first study, we restrict to quasi-circular systems with BH spins parallel or anti-parallel to the orbital angular momentum.  This choice simplifies the joining of separate waveforms,  because  the time shift and spatial rotation
required to align waveforms from two simulations starting at different binary separations
can be determined in post-processing of the simulation data~\cite{Pan:2007nw,Ajith:2007kx}, whereas more
generic systems require a more sophisticated treatment, probably
involving running additional simulations. After validating our approach, we produce three
NR-NR hybrid waveforms for high mass ratios and spin magnitudes, 
and use them to assess the accuracy of two waveform 
models currently used in LIGO and Virgo science,
namely the spinning effective-one-body waveform model calibrated to
NR called {\tt SEOBNRv4}~\cite{Bohe:2016gbl}, and the
inspiral-merger-ringdown phenomenological model called {\tt
IMRPhenomD}~\cite{Khan:2015jqa}.

In the remainder of this article, we describe the \code{ET} and \code{SpEC} codes in more detail, explain and validate our hybridization procedure
to construct NR-NR hybrid waveforms and then compare the newly constructed waveforms to existing waveform models.
Throughout the paper, we use units in which $G = c = 1$.

\begin{table*}
\begin{tabular}{lllllllllllllll}
  \doubleline
  ID & $q^{(S)}$ & $q^{(E)}$ & $\chi_1^{(S)}$ & $\chi_1^{(E)}$ & $\chi_2^{(S)}$ & $\chi_2^{(E)}$ & $e^{(S)}$ & $e^{(E)}$ & $N_\text{orbs}^{(E)}$ & $M_\text{min}^{(S)}/M_\odot$ & $M_\text{min}^{(E)}/M_\odot$\\
\hline
(3,0.85,0.85) & 2.9992 & 2.9986 & 0.850 & 0.851 & 0.849 & 0.851 & $1\times 10^{-4}$ & $2\times 10^{-3}$ & 8.0 & $140.9$ & $279.4$ \\
(3,0.9,0.9) & 2.9988 & 2.9979 & 0.900 & 0.902 & 0.899 & 0.901 & $7\times 10^{-4}$ & $1\times 10^{-3}$ & 8.0 & $180.6$ & $284.5$ \\
(4,0.9,0.9) & 3.9980 & 3.9969 & 0.900 & 0.902 & 0.899 & 0.901 & $9\times 10^{-4}$ & $6\times 10^{-4}$ & 7.5 & $178.3$ & $322.3$ \\
(5,0.9,0.9) & 4.9974 & 4.9955 & 0.900 & 0.902 & 0.899 & 0.901 & $1\times 10^{-3}$ & $8\times 10^{-4}$ & 8.5 & $190.4$ & $321.9$ \\

  \doubleline
\end{tabular}
\caption{Properties of the NR waveforms.  Shown are the case ID,
  the mass ratio $q = m_1/m_2 \ge 1$, the dimensionless spin $\chi_{i}$ of each
  BH ($i=1$ or $2$) in the direction parallel to the orbital angular
  momentum, the eccentricity $e$ near the start of the waveform, the
  number of orbits $N_\text{orbs}$, the minimum masses $M_\text{min}$ for which the
  waveform is entirely in the detector band at $10 \text{Hz}$.
  Superscripts indicate \code{SpEC} (S) and \code{ET} (E).
}
\label{tab:cases}
\end{table*}

{\it Numerical-relativity codes and waveforms.}
\code{SpEC}~\cite{SpECwebsite,Scheel:2006gg,Szilagyi:2009qz,Buchman:2012dw}
is a pseudo-spectral code for efficiently solving partial differential
equations, with the primary goal of modeling compact-object binaries.
\code{SpEC} evolves the first-order formulation~\cite{Lindblom:2005qh}
of Einstein's equations in generalized-harmonic
gauge~\cite{Friedrich1985,Pretorius2005a}
and employs BH excision~\cite{Hemberger:2012jz,Scheel:2014ina}.  For more details see~\cite{Lindblom:2009tu,Pfeiffer:2007yz,Buonanno:2010yk,Szilagyi:2014fna,Boyle:2009vi,Boyle:2013nka,Taylor:2013zia}.

The \code{Einstein Toolkit}~\cite{Loffler:2011ay} is a collection of
open source NR components built around the \code{Cactus}
framework~\cite{Goodale:2002a}.  Our BBH simulations are
based on 8th order finite-differencing and the \textit{moving
  puncture} method to solve the
BSSN~\cite{Nakamura:1987zz,Shibata:1995we,Baumgarte:1998te} or
CCZ4~\cite{Bona:2003fj,Alic:2011gg} formulations of the Einstein
equations for Bowen-York initial data
\cite{Bowen:1980yu,Brandt:1997tf}.  More details of the specific
components and techniques used in the \code{ET}
can be found in~\cite{Ansorg:2004ds,Pfeiffer:2007yz,Brown:2008sb,Baker:2005vv,Campanelli:2005dd,Schnetter:2003rb,Pollney:2009yz,Thornburg:2003sf,Dreyer:2002mx,Reisswig:2010di,Husa:2004ip,Thomas:2010aa,wardell_barry_2016_155394,SimulationToolsWeb}.

\code{SpEC} achieves high accuracy through
pseudo-spectral methods, grids tightly adapted to the shape of the binary and
use of a
computational grid which rotates along with the binary.  The
\code{ET}, in contrast, uses finite-difference methods and
straightforward box-in-box mesh refinement with a global inertial
computational grid.  However, the \code{SpEC} excision algorithm
requires accurate and careful tracking of the BH horizons,
particularly for high-spin BHs,
to preserve purely-outgoing excision boundary
conditions~\cite{Hemberger:2012jz}.  The high accuracy requirement,
combined with strongly deformed BH horizons reduce \code{SpEC}'s
robustness in previously unexplored regions of parameter space,
specifically for higher mass-ratio and simultaneous high spins.  In
contrast, the finite difference moving puncture method (see
Refs.~\cite{Hannam:2006vv,Hannam:2008sg} for a detailed study)
requires no special treatment of the BHs, beyond the choice of
suitable coordinate conditions and the requirement to have sufficient
grid resolution near them.  While finite difference moving puncture
evolutions have proven to be extremely robust, great care is needed in
the choice of grid-structure to achieve high accuracy.

The NR simulations used in this work are detailed in
Table~\ref{tab:cases}, where we label cases with $(q, \chi_{1},
  \chi_{2})$, and where for each case an \code{ET} and a \code{SpEC} simulation is available. At present, we consider only aligned-spin
  simulations, so that $\chi_{1,2}$ represent both the
  spin-projection onto $\mathbf{L}$ as well as the spin-magnitude. The \code{SpEC}
simulation for case (3,0.85,0.85) completed successfully,
  including merger and ringdown.  It is available in the SXS waveform
  catalog~\cite{SXSCatalogue} under ID SXS:BBH:0293 and was first presented in
  Ref. \cite{Chu:2015kft}.
  The remaining \code{SpEC} simulations are new,
start $\sim 20$ orbits before merger, but fail around the merger,
leading to incomplete waveforms.  The \code{ET} simulations are all new;
their length was chosen to be relatively short, but long enough to overlap
with the \code{SpEC}--inspirals.
All simulations are run at
multiple numerical resolutions in order to assess
numerical truncation error.

The \textit{initial data} parameters $(q, \chi_1, \chi_2)$ of the \code{ET} simulations are identical
to those of the \code{SpEC} simulations, but due to uncontrolled emission
of energy and angular momentum from the initial data slice~\cite{Lovelace:2008hd}, the \textit{relaxed} values after the initial dynamical phase
deviate slightly from those of \code{SpEC}.
Table~\ref{tab:cases} presents these relaxed values, which differ
  by $\lesssim 2\times 10^{-3}$ between the two codes.

Some of the \code{ET} evolutions became unstable during
ringdown when evolved with the BSSN formulation.  Presumably, this
  instability is related to the high spin of the remnant BH, 0.93, the
  highest spin BH we have evolved using the \code{ET} code.
  Ref.~\cite{Zlochower:2017bbg} reports significant constraint
  violations in BSSN--evolutions of high spin BHs, which could be
  mitigated by use of the CCZ4 formulation.  Consequently, we re-ran
  the failing simulations with CCZ4 and constraint-damping parameters
  $\kappa_1 = 0.1, \kappa_2 = 0$.  The resulting evolutions are
  stable, but needed higher resolution during the inspiral to maintain
  orbital-phase accuracy.

\newcommand{\hi}{h_\text{insp}}
\newcommand{\hm}{h_\text{merger}}
\newcommand{\hh}{h_\text{hybrid}}
\newcommand{\hS}{h_\text{S}}

\newcommand{\omi}{\omega_\text{insp}}
\newcommand{\omm}{\omega_\text{merger}}
\newcommand{\omh}{\omega_\text{hybrid}}

\newcommand{\Ai}{A_\text{insp}}
\newcommand{\Am}{A_\text{merger}}
\newcommand{\Ah}{A_\text{hybrid}}

\newcommand{\phii}{\phi_\text{insp}}
\newcommand{\phim}{\phi_\text{merger}}
\newcommand{\phih}{\phi_\text{hybrid}}

{\it NR-NR hybrid waveforms and their application.}  We aim to join an inspiral waveform $\hi$ and a merger waveform $\hm$
representing the same astrophysical BBHs, 
where  $\hi$ terminates shortly before the merger peak, and $\hm$
overlaps $\hi$ during the late inspiral.
Because the simulations start at different initial orbital separations,
$\hi$ and $\hm$ will
in general differ by a time translation and a spatial rotation.
We restrict here to a
single spin-weight $-2$ spherical harmonic $\ell=2,m=2$ multipole, and
for notational convenience write $h \equiv h_{22}\equiv A e^{i\phi}$, where the second equivalence defines a decomposition in amplitude and phase. 

To construct the NR-NR hybrid waveform, we must first determine the time and
phase shifts between $\hm$ and $\hi$. We determine 
$\Delta t$ and $\Delta \phi$ from a portion of the waveform in an interval
$[t_1,t_2] = [t_2-T,t_2]$ where $t_2$ is the time
at which $h_i$ terminates and, say $T = 150 M$, where all times are measured in the coordinate
system of $\hi$.
We do not utilize the portion of $\hm$ for $t<t_1$.
For quasi-circular BBH systems with spins aligned with the orbital
angular momentum, the instantaneous GW frequency $\omega\equiv
d\phi/dt$ increases monotonically with time.  We first align $\omega$
in time, which is independent of $\Delta \phi$, and then align the
phases.  All alignments are performed using least-squares
minimisation.  
We conclude the alignment by redefining $\hm(t)\leftarrow e^{i\Delta\phi}\hm(t+\Delta t)$.
We
verified that 
varying the hybridisation interval introduces a negligible error (i.e., 
a mismatch, as defined below, of less than $10^{-4}$).
To achieve a smooth transition between $\hi$ and $\hm$, we
blend $\phii$ with $\phim$ and $\Ai$ with $\Am$
in the interval $[t_1,t_2]$ using a variant of the Planck taper
function~\cite{McKechan:2010kp},
$\mathcal{T}(t;t_1,t_2) = 0$ for $t \le t_1$, $1/\{ \exp [(t_2-t_1)/(t-t_1) + (t_2-t_1)/(t-t_2)] + 1 \}$ for $t_1 < t < t_2$ and $1$ for $t \ge t_2$.
Specifically, we construct $\phih = (1-\alpha) \phii + \alpha \phim\,, 
\Ah = (1-\alpha) \Ai + \alpha \Am$ where $\alpha(t) = \mathcal{T}(t;t_1,t_2)$. Finally, we construct $\hh = \Ah e^{i \phih}$.

\begin{figure}
  \includegraphics[width=0.97\columnwidth]{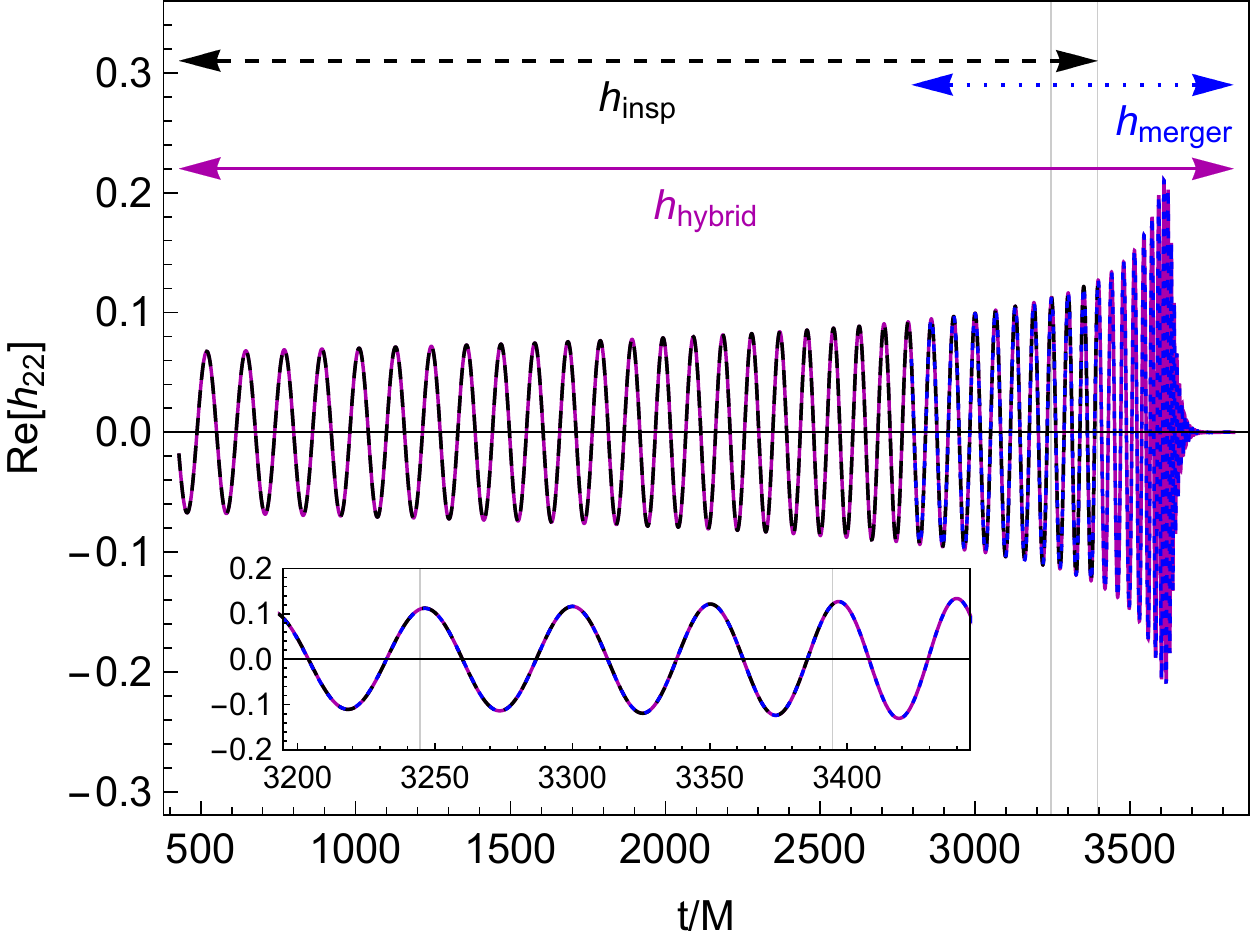}$\quad$
  \caption{Gravitational-wave strain.  Shown is the real part of the
    dominant $\ell=2,m=2$ multipole for the case $(5,0.9,0.9)$ from
    the \code{SpEC} inspiral waveform $\hi$, the ET merger waveform $\hm$,
    and the NR-NR hybrid waveform $\hh$ as a function of retarded time $t$
    in units of the total mass $M$ of the binary.  The vertical lines
    indicate the hybridisation interval $[t_1,t_2]$, which is enlarged in the inset.
  }
  \label{fig:waveforms-aligned}
\end{figure}

Figure~\ref{fig:waveforms-aligned} shows the three waveforms $\hi$,
$\hm$ and $\hh$.  There is no visible discrepancy between the
inspiral, merger and hybrid waveforms in the hybridisation interval.  In
fact, $|\phim-\phii| < 6 \times 10^{-3}$ and $|1-\Am/\Ai| < 4 \times
10^{-4}$ across the hybridisation interval.

NR waveforms are approximations to the solutions of the Einstein
equations, and it is necessary to quantify their uncertainty. 
We consider the impact of
truncation error and extrapolation error on the original inspiral and
merger waveforms.  Truncation error results from using a nonzero grid
spacing in the numerical method.  Extrapolation error arises from the
method used to estimate the asymtpotic GWs (at
future null infinity) from the GWs extracted at finite-radius.
Moreover, an additional error arises because the merger simulation represents a slightly different
physical system than the inspiral simulation (as described above).  We will see that this is
actually the dominant source of error in the NR-NR hybrid waveform.

Gravitational-wave signals from BBHs are often compared using an
optimised detector-noise-weighted inner product.  Given two waveforms
$h_1(t)$ and $h_2(t)$, their noise-weighted \textit{overlap} is
defined as~\cite{Finn:1992wt} $( h_1 | h_2 ) \equiv 4\, \textrm{Re}
\int_{f_{\text{min}}}^{f_{\text{max}}} {\tilde h_1(f) \tilde
  h^*_2(f)}/{ S_n(f) }{\textrm{d}} f$, where $\tilde h_{1,2}(f)$ are
the Fourier transforms of the waveforms and $S_n(f)$ is the one-sided
noise power spectral density, 
  chosen here as the zero-detuned-high-power variant
  from \cite{Shoemaker:2010}.
We use  $f_{\text{min}} = 10\Hz$,
$f_{\text{max}} = 8192 \Hz$.
The \textit{match} between two waveforms is then defined as the
overlap between the normalized waveforms maximized over relative time
and phase shifts, $M(h_1,h_2) \equiv \max_{\phi_c, t_c} \left(h_1 (\phi_c, t_c)~|~h_2\right)/{\sqrt{ (h_1|h_1) (h_2|h_2)}}$. 
The match measures how similar the waveforms appear to a GW detector when the data is analyzed using matched
filtering.  For convenience, since the differences we consider are small, we consider the
\textit{mismatch} $\mathcal{M}(h_1,h_2) \equiv 1 - M(h_1,h_2)$.  A mismatch of zero
indicates that the waveforms appear identical.

\begin{figure}
  \includegraphics[width=0.96\columnwidth]{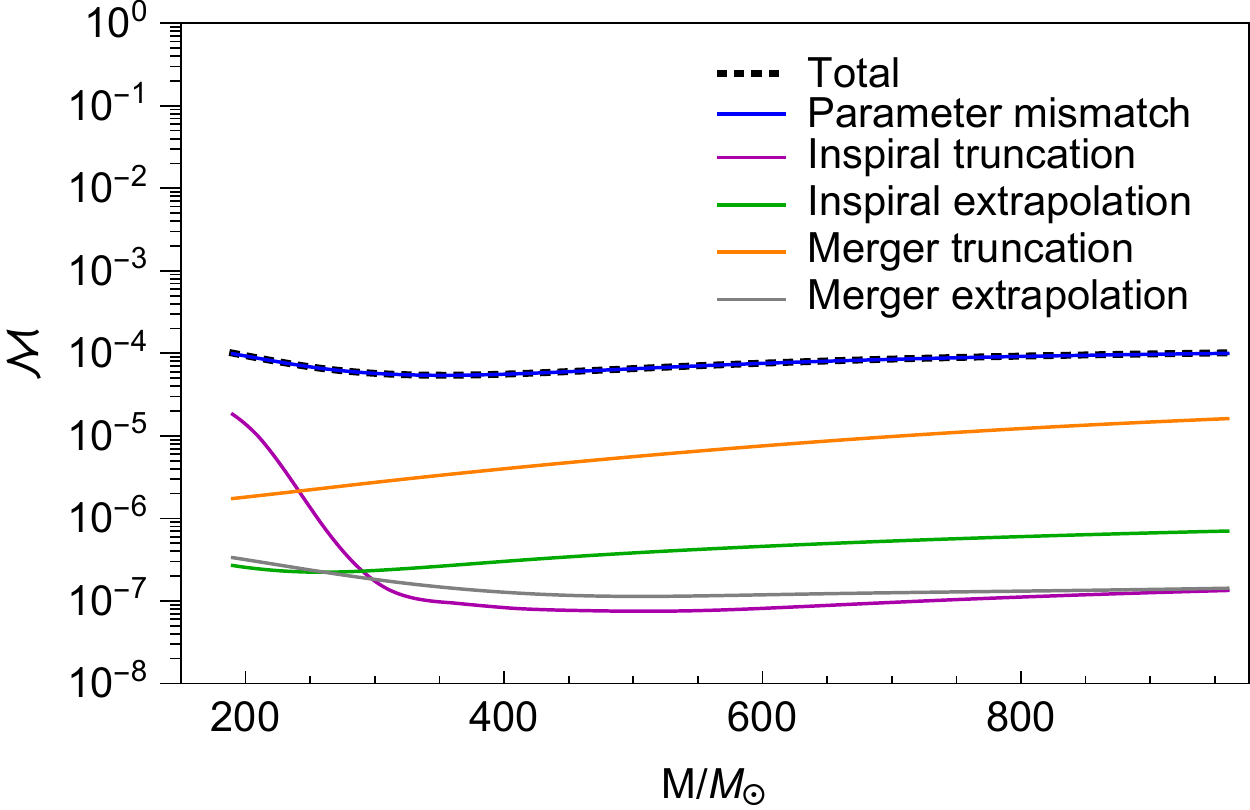}$\quad$
  \caption{
    Estimated uncertainties in the NR-NR hybrid waveform arising from different error sources, expressed in terms of mismatch $\mathcal{M}$ for
    different total binary masses $M$.  Shown is case $(5, 0.9, 0.9)$.
  \label{fig:unfaithfulness-hybrid-errs}
  }
\end{figure}

Figure~\ref{fig:unfaithfulness-hybrid-errs} summarizes the error budget for an NR-NR hybrid for
the case $(5,0.9,0.9)$ as a
function of the total mass $M$ of the binary (a single NR BBH
simulation can represent a binary of any total mass by a
scaling of the waveform).  Each curve represents the mismatch
between a NR-NR hybrid waveform and a \textit{perturbed} NR-NR hybrid.  For the
inspiral and merger truncation error curves, the perturbed hybrid is
constructed using a lower resolution inspiral and merger waveform,
respectively.  Similarly, for the inspiral and merger extrapolation-error 
curves, each perturbed hybrid is constructed using higher-order
extrapolation for the inspiral or merger waveforms, respectively.
Furthermore, the parameter mismatch curve compares a standard {\tt SEOBNRv4} waveform~\cite{Bohe:2016gbl}
with parameters of 
the inspiral simulation with a hybrid waveform constructed 
in the same way as the NR-NR hybrid but from two {\tt SEOBNRv4} waveforms, the first using the parameters of the inspiral
simulation, and the second using the parameters of the merger
simulation.  We assume that the mismatch introduced by
hybridising two {\tt SEOBNRv4} waveforms with different parameters is
representative of the same effect with NR waveforms.

The total error budget in Fig.~\ref{fig:unfaithfulness-hybrid-errs} is the quadrature-sum of the individual contributions, and provides a
conservative indication of the total uncertainty in the waveform as
seen by a GW detector.
Figure~\ref{fig:unfaithfulness-hybrid-errs} demonstrates that the 
error of the hybrid waveform is dominated by the slightly inconsistent
BBH parameters for $\hi$ and $\hm$.
  This is also the case for all the other
  NR configurations.  These parameter inconsistencies
could be reduced by
repeating the merger simulation with initial mass ratios and spins
adjusted in such a way that the relaxed values approach those of the
inspiral simulation.  In the future, this step should be performed at the same time as
the iterative eccentricity reduction, which already requires multiple simulations.

\begin{figure}
  \includegraphics[width=0.95\columnwidth]{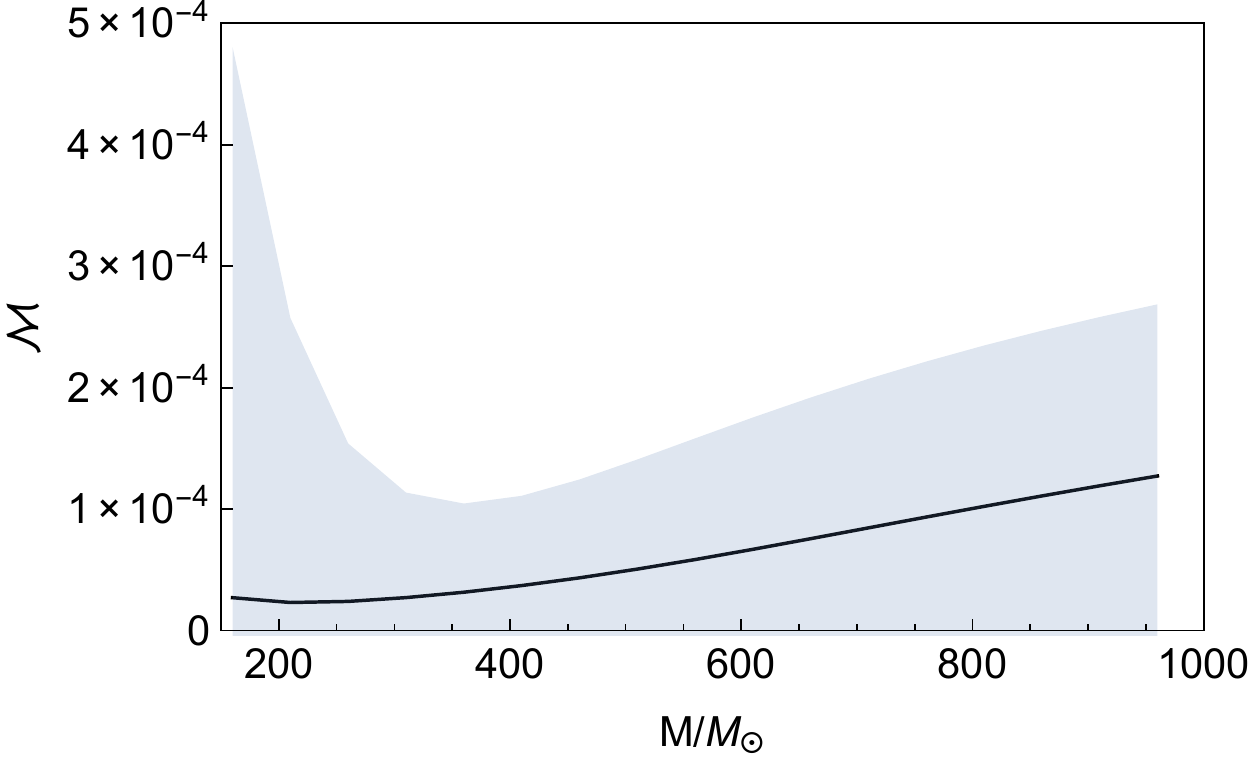}$\quad$
  \caption{Mismatch $\mathcal{M}$ between SpEC-ET hybrid and complete SpEC
    waveforms for the case (3,0.85,0.85) where the \code{SpEC} waveform
    includes a merger.  The shaded region represents the estimated
    uncertainty in this computation coming from errors in the \code{SpEC} and
    \code{ET} waveforms.}
  \label{fig:full-compare-uf}
\end{figure}

Figure~\ref{fig:full-compare-uf} tests our hybridisation method
and our error estimates for the case $(3, 0.85, 0.85)$, for which
complete inspiral-merger-ringdown \code{SpEC}--waveforms are available.
Discarding the \code{SpEC} merger, we construct a
  \code{SpEC}--\code{ET} hybrid waveform as described above, and plot its mismatch with the complete \code{SpEC} waveform as the solid black line.  The grey band in Fig.~\ref{fig:full-compare-uf} represents our error-estimate on the hybrid, computed as the quadrature sum of the effects on the
mismatch between $\hh$ and $\hi$ of perturbing the input
waveforms by their individual sources of error.   Figure~\ref{fig:unfaithfulness-hybrid-errs} indicates that the hybridization introduces only small errors ($\mathcal{M}\lesssim 10^{-4}$), and confirms the validity of our error estimate from Fig.~\ref{fig:unfaithfulness-hybrid-errs}.

\begin{table}[b]
\begin{tabular}{lllllllllllllll}
  \doubleline
  ID & $N_\text{orbs}^{(H)}$ & $\mathcal{M}_\text{hybrid}^\delta$\\
\hline
(3,0.85,0.85) & 23.9 & $3\times 10^{-5}$ \\
(3,0.9,0.9) & 16.3 & $7\times 10^{-5}$ \\
(4,0.9,0.9) & 19.3 & $1\times 10^{-4}$ \\
(5,0.9,0.9) & 20.0 & $1\times 10^{-4}$ \\

  \doubleline
\end{tabular}
\caption{Properties of the NR-NR hybrid waveforms.  Shown are the case
  ID, the number of orbits $N_\text{orbs}$,
  and the estimated
  uncertainty $\mathcal{M}_\text{hybrid}^\delta$ on the hybrid, computed as
  described in the text.}
\label{tab:results}
\end{table}

We
construct NR-NR hybrid waveforms for each of the configurations
shown in Table~\ref{tab:cases}, with properties of these NR-NR hybrids given in Table~\ref{tab:results}.
All cases have errors comparable to
the validation case $(3,0.85,0.85)$ shown in
Fig.~\ref{fig:unfaithfulness-hybrid-errs}.  In each case, the dominant
source of error is the parameter mismatch between the inspiral and
merger simulations.  The maximum estimated uncertainty in the NR-NR hybrid
across all the configurations is a mismatch of $\hybridmaxuferr{}$.

It is not in general known how the mismatch between an approximate waveform and
the true waveform relates to biases in parameter recovery except in specific
configurations which have been studied. Ref.~\cite{Abbott:2016wiq} found that a
waveform mismatch of $\sim 5\times 10^{-3}$ did not lead to noticeable parameter
estimation biases for binaries comparable to GW150914\footnote{Except for nearly
  edge-on systems where errors are dominated by the lack of sub-dominant modes
  in the waveform models.}, though it is not clear how to translate this result
to the more extreme configurations we study here.
Moreover, differences between the analytical waveform models
  considered in the next section and our NR-NR hybrids are $\sim
  10$-times larger than ${\mathcal M}^\delta_{\rm hybrid}$ reported in
    Table~\ref{tab:results}.
The waveform models SEOBNRv4
and IMRPhenomD were found to be accurate to NR waveforms to a mismatch of
$\mathcal{M} \lesssim 10^{-2}$, so the NR-NR hybrid waveforms are an order of
magnitude more accurate, and hence suitable for comparing with these waveform
models.
Finally, the distinguishability criterion
\cite{Flanagan:1997kp,Lindblom:2008cm,McWilliams:2010eq,Chatziioannou:2017tdw},
$\mathcal{M}<D/(2\rho^{2})$ where $\rho$ is the SNR of the signal and $D$ is
  the number of intrinsic parameters that would not be measured accurately due
  to inaccuracy of the waveform model.  For aligned-spin systems considered in this
  paper we take $D=4$, and using the largest mismatch from all the
  configurations, we find that the waveforms presented here would be
  indistinguishable by Advanced LIGO for signal-to-noise ratios $\lesssim 140$,
which includes most of the expected detectable events. We are therefore
satisfied that the NR-NR hybrid waveforms are sufficiently accurate for use in
waveform modelling.

As a first application of our synergistic approach of combining NR codes to 
compute waveforms for large mass ratios and spins,
we compare the new NR-NR hybrid waveforms with
the analytical 
waveform models {\tt SEOBNRv4} and {\tt IMRPhenomD} currently employed
in LIGO and Virgo science.
We evaluate these waveform models at the relaxed values of
mass ratio and spin computed from the \code{SpEC} simulations
in Table~\ref{tab:cases}.

\begin{figure}
  \includegraphics[width=0.95\columnwidth]{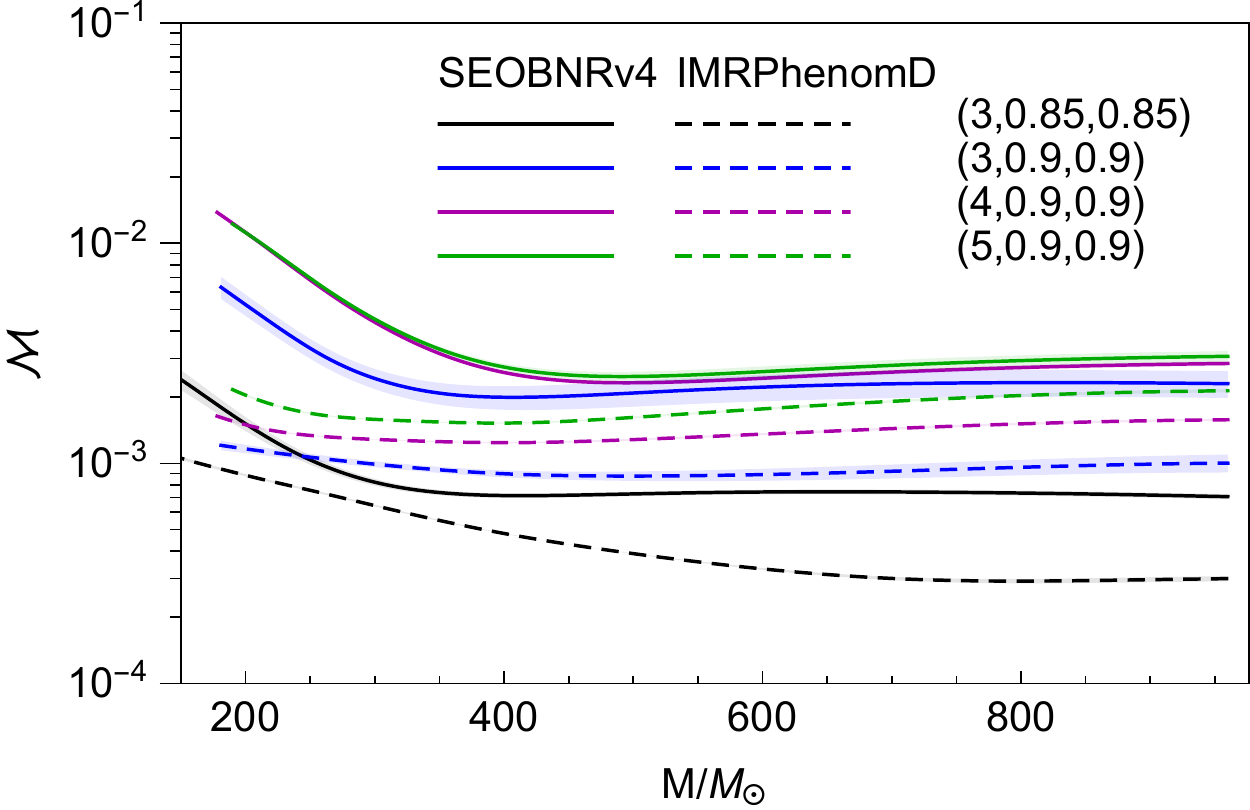}$\quad$
    \caption{Mismatch $\mathcal{M}$ between SpEC-ET hybrids and the waveform
    models {\tt SEOBNRv4} (solid lines) and {\tt IMRPhenomD} (dashed lines) for
    all the configurations in Table \ref{tab:cases}.}
  \label{fig:hybrid-eob-uf}
\end{figure}

Figure~\ref{fig:hybrid-eob-uf} shows the mismatch between the
NR-NR hybrid and the waveform models for each case.  The very thin, shaded regions around each curve 
indicate the estimated error of the NR-NR hybrid, computed as the quadrature sum of the effects on the
mismatch between $\hh$ and the model waveforms of perturbing the $\hh$
by their individual sources of error.   These errors in the hybrids themselves are negligible in comparison with the difference between the
NR-NR hybrids and the waveform models.  The curves start at the
minimum masses for which the NR waveforms reach down to 10 Hz, and
hence cover the frequency range of Advanced LIGO and Virgo at design
sensitivity.  Comparisons would only be possible at significantly higher total mass
if only the shorter \code{ET} waveforms were used.

The waveform models had mismatches of typically $\lesssim 10^{-2}$ with the
NR waveforms they were originally tested (or calibrated) against.  We see that for the more
extreme configurations presented here, both waveform models maintain this
accuracy.
The mass-range $M\gtrsim 200 M_\odot$ of the comparison in Fig.~\ref{fig:hybrid-eob-uf} is determined by the requirement that the NR-NR hybrid waveforms cover the entire Advanced LIGO sensitivity band.
Waveforms with lower initial orbital frequency (i.e.~more
orbits) are required to extend this calculation to lower mass.
The NR waveforms produced here could
now be used to improve the waveform models, if more accuracy in this region of
the parameter space is required, for example for future GW detectors with higher
sensitivities. 


{\it Conclusions.} BBH simulations for systems with both high mass ratio
and high spin remain challenging for NR codes, and yet it is in this
region of the parameter space where waveform models typically depend
strongly on calibration to NR simulations~\cite{Khan:2015jqa,Bohe:2016gbl}. 
The \code{SpEC} code, based on accurate pseudo-spectral methods, sophisticated adapted grids and
corotating coordinates has a long history of producing large numbers of long ($\gtrsim 25$ orbits)
simulations (e.g., see Refs.~\cite{Mroue:2013xna,Chu:2015kft}), but sometimes requires significant fine-tuning in order
to simulate the merger \cite{Scheel:2008rj,Scheel:2014ina}.  Finite difference codes,
on the other hand, have generally been used to produce shorter
simulations, typically $\lesssim 15$ orbits (see, e.g., Refs.~\cite{Healy:2017psd,Jani:2016wkt, Husa:2015iqa}),
possibly reflecting a lower accuracy for a given
computational cost, due to the use of more straightforward numerical
methods.  However, maybe as a result of this simplicity, these
codes are able to handle some more extreme BBH configurations where
SpEC currently fails to simulate the merger.

We have taken advantage of the strengths of both codes to combine
accurate and efficient \code{SpEC} inspirals with robust finite difference
mergers, and have expanded the parameter space of available NR
waveforms.  With the resulting new NR hybrid waveforms, we test
the state-of-the-art waveform models {\tt SEOBNRv4} and {\tt IMRPhenomD}.
We find that the models agree with the
new NR waveforms to mismatch $\mathcal{M} \lesssim 1\%$, for total
binary masses for which the entire NR-NR hybrid waveform is in the Advanced LIGO
detector band starting at 10 Hz.  This is sufficient accuracy for
current waveform modelling purposes.

While we anticipate that the problems encountered at the merger with
SpEC will be resolved with improved numerical methods, our
synergistic hybridisation approach provides a useful stop-gap measure to extend
the science reach of current NR codes, and hence the parameter space
of waveforms available for waveform-model validation for upcoming
GW science with LIGO and Virgo detectors.  The method is completely 
generic for aligned-spin binaries and is
applicable to any pair of NR codes.  A useful next step would be to apply the method
to the subdominant modes ($\ell > 2$), as these are important for systems with the high mass ratios studied here. For precessing binaries, however, it requires techniques to match the precessing spin-directions between the two waveforms at a certain reference frequency.

All the waveforms presented here, including the NR-NR hybrids, are
publicly available~\cite{hinder_ian_2018_1469114}.  An implementation
of the hybrid construction method is available in the SimulationTools
for Mathematica~\cite{SimulationToolsWeb} package, and a Mathematica
notebook demonstrating the method is provided in
\cite{hinder_ian_2018_1469114}.

{\it Acknowledgments.} Calculations were performed using the Spectral Einstein code ({\tt
  SpEC})~\cite{SpECwebsite} and the Einstein Toolkit~\cite{etweb} on
the {\tt Minerva} high-performance computer cluster at the Max Planck Institute for Gravitational Physics 
in Potsdam. The Einstein Toolkit simulations were originally based on the Einstein
Toolkit GW150914 BBH example \cite{wardell_barry_2016_155394}.
Analysis of the numerical data was performed using SimulationTools for
Mathematica~\cite{SimulationToolsWeb}.
H.P. gratefully acknowledges funding from NSERC of Canada, the Ontario
Early Researcher Awards Program, and the Canada Research Chairs Program.

\bibliography{references}

\end{document}